\documentstyle[12pt,a4,epsf]{article}
\begin{document}
\def\IR{\relax{\rm I\kern-.18em R}}
\font\cmss=cmss12 \font\cmsss=cmss12 at 7pt
\def\IZ{\relax\ifmmode\mathchoice
{\hbox{\cmss Z\kern-.4em Z}}{\hbox{\cmss Z\kern-.4em Z}}
{\lower.9pt\hbox{\cmsss Z\kern-.4em Z}}
{\lower1.2pt\hbox{\cmsss Z\kern-.4em Z}}\else{\cmss Z\kern-.4em Z}\fi}
\def\pla{\alpha'}
\baselineskip18pt


\thispagestyle{empty}

\begin{flushright}
\begin{tabular}{l}
FFUOV-97/02\\
{\tt hep-th/9705185}\\
\end{tabular}
\end{flushright}

\vspace*{2cm}

{\vbox{\centerline{{\Large{\bf COUNTING CLOSED STRING STATES
}}}}}

\vspace{0.8cm}

{\vbox{\centerline{{\Large{\bf 
IN A BOX }}}}}

\vskip30pt

\centerline{Marco Laucelli Meana, M.A.R. Osorio, and Jes\'{u}s Puente Pe\~{n}alba 
\footnote{E-mail address:
     laucelli, osorio, jesus@string1.ciencias.uniovi.es}}

\vskip6pt
\centerline{{\it Dpto. de F\'{\i}sica, Universidad de Oviedo}}
\centerline{{\it Avda. Calvo Sotelo 18}}
\centerline{{\it E-33007 Oviedo, Asturias, Spain}}

\vskip .25in

\vspace{1cm}
{\vbox{\centerline{{\bf ABSTRACT }}}}

The computation of the microcanonical density of states for a string gas in a finite volume needs a one by one count because of the discrete nature of the spectrum. We present a way to do it using geometrical arguments in phase space. We take advantage of this result in order to obtain the thermodynamical magnitudes of the system. We show that the results for an open universe  exactly coincide with the infinite volume limit of the expression obtained for the gas in a box. For any finite volume the Hagedorn temperature is a maximum one, and the specific heat is always positive. 
We also present a definition of pressure compatible with $R$-duality seen as an exact symmetry, which allows us to make a study on the physical phase space of the system. Besides a maximum temperature the gas presents an asymptotic pressure. 
 
\vspace*{24pt}

\setcounter{page}{0}
\setcounter{footnote}{0}

\newpage

\section{Introduction}

Now that we know at last that, in the infinite volume limit, the microcanonical description of a gas of closed strings undergoes, at a finite energy per string, a transition to a phase with a maximum temperature evolving from increasing positive specific heat \cite{us}, it seems natural to explore a more physical way of achieving big (infinite) volumes by departing from small ones.

Historically this is not a new topic. Many times in Physics, questions do not arise as a consequence of logical focuses; but from rather biased assumptions.
It is in the context of String Cosmology where the question of stringy universes contained in boxes raised the problem of taking the infinite volume limit in relation with the volume dependence of the microcanonical density of states \cite{BV}. Calculations for the high energy limit of the system yielded a picture in which the infinite volume limit taken from finite volumes in the microcanonical ensemble produced a scenario different from the one obtained by getting the microcanonical description directly from already open universes \cite{DJT}. This astonishing result was explained as a pure stringy effect stemming from the string windings.

It is a little bit surprising that such conclusion can be supported by a calculation that is only valid for a vague high energy of the system.  What we propose, because it is feasible, is to compute the complete density of states of the single string valid for any value of the energy and for any volume. When one has discrete momenta and, of course, discrete windings one should count their contribution to the degeneration one by one. This is actually what we are going to do in section three after some introduction to the microcanonical description in String Statistical Mechanics. In the fourth section, we will take the infinite volume limit and we will analytically show that we exactly recover the result that is found by inverse Laplace transforming the infinite volume free energy computed through the $S$-representation. We will also show that the same happens for the multiple string gas. Finally, we present a study of the behaviour of the system in terms of its volume, focusing on the pressure and the system's phase diagram. 

We will choose closed superstrings, but not the Heterotic, to work with. However it is not hard to see that all the conclusions hold for any closed string  gas, and also for the Heterotic one.

\section{A reminder on the Microcanonical Ensemble}

Let us begin with a brief review about the forms of defining the microcanonical density of states (see \cite{huang}, for example). Suppose we have a system with a continuous spectrum of energies, then we can call $\Gamma(E,V)$  the number of accesible states falling into an interval $[E,E+dE]$. The density of states is defined by:

\begin{equation}
\Gamma(E,V)=\Omega(E,V)dE
\label{omega}
\end{equation}   

If the system has got a discrete spectrum of energies, the differential of $E$ cannot be taken arbitrarily small, but has a minimum value.
In this situation what seems to us as the most fundamental way to compute $\Gamma(E,V)$ is to enumerate all the states in which the system can appear at a given energy interval.

The definition of the partition function of a system is 
\begin{equation}
Z(\beta,V)=\sum_{\{n\}} e^{-\beta E(\{n\})}
\end{equation}
where $\{n\}$ represents all the quantum numbers defining the state and its  energy  is $E(\{n\})$.  It will also be  useful to recall that for  a string gas  that has null chemical potential, there are contributions from  all the possible $N$-string systems, and the partition function of each one is
\begin{equation}
z^{(N)}(\beta,V)=\frac{1}{N!}\left(-\beta F(\beta,V)\right)^N
\end{equation}
Now the partition function can be written as 
\begin{equation}
 Z(\beta,V) = \sum_{N=0}^{+\infty}z^{(N)} = \exp\left[{-\beta F(\beta,V)}\right]
\end{equation}
We must note that there is a contribution from the system with no subsystems, i.e. $N=0$; quantum mechanically  this is the vacuum at the temperature of the bath. This term makes the main difference when comparing the canonical and microcanonical ensembles. When the energy exchange between the vacuum and the bath is relevant one should not forget that if both ensembles describe systems with the same entropy, then magnitudes like the specific heat may not coincide when expressed in terms of $E$ (microcanonical ensemble) or $T$ (canonical ensemble). 

We will focus now on the single string system. With $N=1$ we have
\begin{equation}
z^{(1)}(\beta,V)=- \beta F(\beta,V)=\sum_{i} \Gamma_1(E_i,V) e^{-\beta E_i}
\label{single}
\end{equation}
where $\Gamma_1(E_i,V)$ amounts to the total degeneration for a given energy. The single string density of states contains all the information that suffices to describe the complete system because it is possible to recover any $N$-string system density of states from it. For example, if one has a system composed by two subsystems with densities $\Omega(E,V)$ and $\Omega'(E,V)$, the density of states of the whole system will be 
\begin{equation}
\Omega_{total}(E,V)={\cal A}\int_0^E dt \,\Omega(t,V)\, \Omega'(E-t,V) 
\end{equation}
where $E$ is the total energy shared by both subsystems. The meaning of this integral is obvious, it is the summation over all possible energy distributions. ${\cal A}$ is a combinatorial factor which takes into account the indistinguishability features  of the subsystems. Seen from the point of view of the inverse Laplace tranform techniques the above equation corresponds to the convolution theorem.
  
On the other hand, for the Helmholtz free energy we have:
\begin{equation}
-\beta F(\beta,V)_{B,F}=\pm \sum_{\{n\}}\log{\left(1 \mp e^{-\beta E(\{n\})}\right)}=\mp \sum_{\{n\}} \sum^{\infty}_{r=1}\frac{(\pm 1)^{r}}{r}e^{-\beta r E(\{n\})}
\label{helmholtz}
\end{equation}
where $B (F)$ stands for bosons (fermions). It is important to note that in the boson case  the free energy is divergent  at every temperature if the zero energy state is taken into account. This is the result of counting the vacuum state --with zero energy, and correspondingly all its quantum numbers null-- an infinite number of times if one tries to second quantize it.  The solution is simply to neglect the vacuum contribution. This does not affect the thermodynamical properties in the microcanonical ensemble because there is only one state at zero energy. The quantum corrections to the Maxwell-Boltzmann statistics are represented by the $r>1$ terms in the previous expression. Equating  eq.(\ref{helmholtz}) and eq.(\ref{single}) one sees that $\Gamma_1(E_i,V)$ must take into account quantum statistical effects.

\section{The single string density of states}

The mass formula for a non heterotic closed superstring living in a nine-dimensional hypertorus is \cite{schwarz}

\begin{equation}
\frac{\alpha'm^2}{2}=\frac{1}{2} \frac{R^2}{\alpha'} \vec{m}^2 + 
\frac{1}{2} \frac{\alpha'}{R^2} \vec{n}^{\,2} + N + \tilde{N}
\end{equation}

Where $R$ is the radius of every spatial dimension of the universe, $\vec{n}$ is the quantum number related to the momenta and $\vec{m}$ is the one related to the windings. Both are vectors of the lattice $\IZ^9$. $\tilde{N}$ and $N$ are the total number of left and right oscillators. There also exists the constraint

\begin{equation}
N-\tilde{N}+\vec{m}\cdot\vec{n}=0
\label{liga}
\end{equation}

 To our purpose it is useful to introduce the constraint into the mass formula taking $\tilde{N}=N+\vec{m}\cdot\vec{n}\geq 0$, and put the part corresponding to the windings and discrete momenta in the form of a perfect square:

\begin{equation}
\alpha'm^2=\left(\frac{R}{\sqrt{\alpha'}} \vec{m} + 
\frac{\sqrt{\alpha'}}{R} \vec{n}\right) ^2 + 4 N
\label{mass}
\end{equation}

Now if one notices that because of the finite volume this expression for the mass is in fact the energy of the string, counting the number of ways of getting a given value of $m^2$ corresponds to computing the microcanonical density of states. For the time being, we are going to compute this density for Maxwell-Boltzmann statistics (the $r=1$ term in eq.(\ref{helmholtz})) and we will introduce the quantum corrections later.

To proceed, let us fix some notation. For a while, we will use new variables by setting

\begin{eqnarray}
L = \frac{R}{\sqrt{\pla}}\\
\vec{s} = \frac{1}{L}\, \vec{n} + L\,\vec{m}
\end{eqnarray}

The meaning of $L$ is obvious. That of $\vec{s}$ is a little bit subtle; in some sense, this vector represents the kinetic degrees of freedom (windings and momenta). In terms of these, the constraint appears as

\begin{equation}
\tilde{N} = N + \vec{n}\cdot\left(\frac{1}{L}\,\vec{s} - \frac{1}{L^2}\,\vec{n}\right) = 
N + \frac{1}{L}\,\vec{n}\cdot\vec{s} - \frac{1}{L^2}\,{\vec{n}}^{\,2}
\label{parabol} 
\end{equation}

\begin{figure}
\let\picnaturalsize=N
\def\picsize{2in}
\def\picfilename{parabol.eps}
\ifx\nopictures Y\else{\ifx\epsfloaded Y\else\input epsf \fi
\let\epsfloaded=Y
\centerline{\ifx\picnaturalsize N\epsfxsize \picsize\fi
\epsfbox{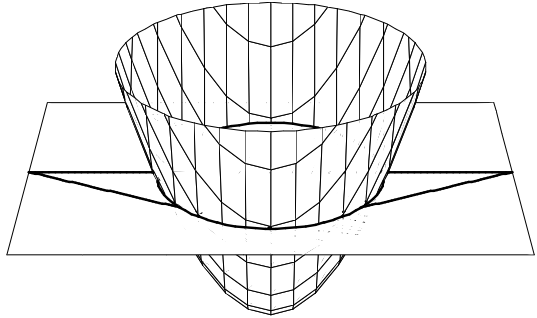}}}\fi
\caption{\small {A three  dimensional version of the geometry involved in the computation of $\Gamma_1(E,V)$.}}
\label{parabola}
\end{figure}

Let us now  keep $N$ and $\tilde{N}$ fixed so that, obeying eq.(\ref{mass}),  $\vec{s}^{\,\,2}$  will be a constant for a given fixed energy. In addition to this we completely fix the vector $\vec{s}$, not only its modulus but also its direction. By doing this, eq.(\ref{parabol}) becomes geometrically the equation of the eight-dimensional hypersphere defined by the intersection of the hyperparaboloid defined by the right hand side of eq.(\ref{parabol}) and the hyperplane defined by making $\tilde{N}$ constant (Fig. 1 may be of some help). All these geometrical objects live in phase space. Geometry will serve our purpose of counting states because we are going to use the volume given by the area of the hypersphere times a differential of its radius  $\rho$ as a continuous measure of the number of discrete states. These states have an energy in an interval whose width is determined by the finite increment of $\rho$. The increment allows us to relate $\Omega_1(E,V)$ with this calculation of $\Gamma_1(E,V)$ as it has been stated in eq.(\ref{omega}).

The sphere is generated by letting the tip of the vector $\vec{n}$ (or equivalently $\vec{m}$) rotate around the center, and then, for the given value of $\vec{s}$ this motion produces a vector $\vec{m}$ ($\vec{n}$) set by  

\begin{eqnarray}
\vec{m} & = & \frac{1}{L} \vec{s} - \frac{1}{L^2} \vec{n} \nonumber \\ 
\left(\vec{n}\right. & = & \left. L \vec{s} - L^2 \vec{m} \right)
\label{vectors}
\end{eqnarray}

For arbitrary $E$, $N$ and $\tilde{N}$, the last equation can violate the fact that $\vec{m}\,(\vec{n})$ is an integer vector. This does not worry us because the state with $\vec{m}\,(\vec{n})$ being the nearest integer vector to the one defined by the relation in eq.(\ref{vectors}) will be in the shell $[\rho,\rho+\Delta\rho]$ if we set $\Delta\rho = 1$, which is the length, in phase space, from a state to the next along the edges of the box. 
We have to decide whether we let $\vec{m}$ or 
$\vec{n}$ rotate. The election is in fact arbitrary but if we want our $\Delta\rho$ to produce the smallest $\Delta E$, we should better choose $\vec{n}$ if $L<1$ and $\vec{m}$ otherwise.
With other elections both $\Gamma_1(E,V)$ and $\Delta E$ would be larger, leading to the same result for $\Omega_1(E,V)$.

The radius of the hypersphere is easily seen to be 
\begin{equation}
\rho=\left(L\,\vartheta(1-L)+\frac{1}{L}\, \vartheta(L-1)\right)\,\sqrt{N-\tilde{N}+\frac{1}{4}\vec{s}^{\,2}} 
\label{rho}
\end{equation}
where $\vartheta(x)$ is the Heaviside function, that comes up to the stage to help us to  express our 'double' election of $\Delta\rho$. This choice is valid provided the assumption that none of the kinetic vectors is null, because in that case everything would vanish. We shall consider those possibilities later on. 

The area of a sphere of radius $\rho$ in $d$-dimensions  is\footnote{$A(d,\rho)$ as a function of $d$; i.e. the number of compactified dimensions, has a maximum
for $d\simeq 7$. Notice that $11 = 7 + 4$}:
\begin{equation}
A(d,\rho)=\frac{2\pi^{d/2}}{\Gamma(d/2)} \rho^{d-1}
\end{equation}

 In eq.(\ref{parabol}) we had fixed the vector $\vec{s}$ to a constant. We have then to let it rotate generating another sphere with radius $\left(L\,\vartheta(1-L)+\frac{1}{L}\,\vartheta(L-1)\right)\,|\vec{s}|$,
where the Heaviside  functions appear for the same reasons they did in eq.(\ref{rho}) above.

After that we have got the degeneration for $N$ and $ \tilde{N}$ fixed such that

\begin{eqnarray}
\Gamma_{N,\tilde{N}}= \frac{1}{2}\,2^8\,a_N\,a_{\tilde{N}}\left(\frac{2\pi^{9/2}}{\Gamma(9/2)}\right)^2 \left|\vec{s}\right|^8\,
\left(\sqrt{N-\tilde{N}+\frac{1}{4}\vec{s}^{\,\,2}}\right)^8 \nonumber\\
\times \left(L\,\vartheta(1-L)+\frac{1}{L}\, \vartheta(L-1)\right)^{16}\Delta\rho
\label{gnn}
\end{eqnarray}

 The factor one half in front of eq.(\ref{gnn}) comes from the fact that we are counting only either bosons or fermions (the theory is supersymmetric) and the factor two to the eighth counts the fermion zero modes.The $a_N$ and $a_{\tilde{N}}$ measure the degeneration tied to the oscillators that give the same $N$ and $\tilde{N}$. They are gotten as the coefficients of the series expansion on $x=e^{2\pi i \tau}$ of the SSTII partition function. They are given by:

\begin{equation}
a_k=\frac{1}{k!}\,\frac{d^k}{dx^k}\left[\theta_4(0,x)^{-8}\right]_{x=0}
\end{equation}

If we substitute $\vec{s}^{\,2} = \pla E^2 - 4N$ in eq.(\ref{gnn}), and sum over all possible values of $N,\tilde{N}$ we obtain
\begin{eqnarray}
\Gamma'(E,V)=\frac{1}{2}\sum_{N=0}^{[\frac{\pla E^2}{4}]}
 a_N \left(\frac{2\pi^{9/2}}{\Gamma(9/2)}\right) (\pla E^2-4N)^{8/2}
 \left(L\,\vartheta(1-L)+\frac{1}{L}\, \vartheta(L-1)\right)^{8}  \nonumber \\
\times \sum_{\tilde{N}=0}^{[\frac{\pla E^2}{4}]}a_{\tilde{N}}
\left(\frac{2\pi^{9/2}}{\Gamma(9/2)}\right)(\pla E^2-4\tilde{N})^{8/2} 
\left(L\,\vartheta(1-L)+\frac{1}{L}\, \vartheta(L-1)\right)^{8} \Delta\rho 
\hspace{1cm}
\label{gammap}
\end{eqnarray}

This is the expression for the degeneration of the single string not forgetting that we have not included the states for which the winding or momentum are null. 
In these cases the hypersphere of radius $\rho$ in phase space must be substituted by one single point, i.e. one single state. We have just one sphere with radius $\sqrt{\pla E^2 - 4\,N}$ multiplied either by $L$ if there are no windings or by $1/L$ if no momenta appear.

Summing everything up we finally arrive to:

\begin{eqnarray}
\Gamma_1(E,V)  = \left[\frac{1}{2}\sum_{N=0}^{[\frac{\pla E^2}{4}]}
 a_N \left(\frac{2\pi^{9/2}}{\Gamma(9/2)}\right) (\pla E^2-4N)^{8/2}
 \left(L\,\vartheta(1-L)+\frac{1}{L}\, \vartheta(L-1)\right)^{8}\right.  \nonumber \\
\left.\times \sum_{\tilde{N}=0}^{[\frac{\pla E^2}{4}]}a_{\tilde{N}}
\left(\frac{2\pi^{9/2}}{\Gamma(9/2)}\right)(\pla E^2-4\tilde{N})^{8/2} 
\left(L\,\vartheta(1-L)+\frac{1}{L}\, \vartheta(L-1)\right)^{8}\right. \nonumber\\
\left.+ \frac{1}{2}\,2^8\,\sum_{k=0}^{[\frac{\pla E^2}{4}]}
 a_k^2 \left(\frac{2\pi^{9/2}}{\Gamma(9/2)}\right) (\pla E^2-4k)^{8/2}
 \left( L^8 + \frac{1}{L^8}\right) \right]\, \Delta\rho  \nonumber 
\\  +  \frac{1}{2}2^8 \sum_{k=1}^{\infty} a_k^2 \delta\left(\left[\frac{\pla E^2}{4}\right],k\right)\hspace{2mm}
\Delta k \hspace{1cm}
\label{total}
\end{eqnarray}
Here $[\frac{\pla E^2}{4}]$ stands for the integer part of what is inside the square brackets. This is so because we have taken the energy as a continuous variable although it actually takes only discrete values. This is a good approximation except for the low energy range of a string in a very little (approximately selfdual) box. The last term comes from the marginal case $\vec{n} = \vec{m} = \vec{0}$. 
Because of the Kronecker deltas this term contributes with a finite number of states only when the energy lies in between two mass levels. Here, we can easily introduce quantum statistics by simply making the same calculation with $rE$ instead of $E$ and multiplying by $1/r$ as can be seen in eq.(\ref{helmholtz}). Adding up the contributions of fermions and bosons reduces the sum over $r$ to the odd values.

We should not miss the point that our goal is to get the density of states $\Omega_1(E,V)$. So, $\Gamma_1(E,V)$ must be devided by $\Delta E$. Taking into account that 

\begin{equation}
\frac{\Delta\rho}{\Delta E} = \frac{\Delta\rho}{\Delta|\vec{s}|}
\frac{\Delta|\vec{s}|}{\Delta E} = \frac{\pla E/r^2}{\sqrt{\pla E^2/r^2 - 4N}}
\frac{\Delta\rho}{\Delta|\vec{s}|}
\end{equation}
where $\Delta\rho/\Delta|\vec{s}|$ equals either $L$ or $1/L$ depending on the case (cf.  for example eq.(\ref{rho})), and
\begin{equation}
\frac{\Delta k}{\Delta E}=\frac{\pla  E}{2r^2}
\end{equation}

At this point, we are able to write a closed expression for the density of states of the single string valid across the whole range of energy and volume

\begin{eqnarray}
\Omega_1(E,V)=
\sum_{(r\geq 1,\, odd)}
\sum_{N,\tilde{N}=0}^{\infty}
\frac{a_N a_{\tilde{N}}}{r^{18}}\left(\frac{2\pi^{9/2}}{\Gamma(9/2)}\right)^2
\vartheta\left(E^2-\frac{4 r^2 N}{\alpha'}\right) \vartheta\left(E^2-\frac{4 r^2 \tilde{N}}{\alpha'}\right)
 \nonumber\\  \times \,
\pla E \left(\pla E^2-4Nr^2\right)^{7/2}\left(\pla E^2-4\tilde{N}r^2\right)^{8/2}\Lambda(R) \nonumber \\
%
%
%
+\sum_{(r\geq 1\,,odd)}
\sum_{k=0}^{\infty}\frac{2^8a_k^2}{r^{10}}
 \left(\frac{2\pi^{9/2}}{\Gamma(9/2)}\right) \pla\,E \left(\pla E^2-4k
r^2\right)^{7/2}\vartheta\left(E^2-\frac{4 r^2 k}{\alpha'}\right)\Upsilon(R)\nonumber\\
%
%
+ 2^8 \sum_{(r\geq 1\,,odd)}\sum_{k=1}^{\infty} a_k^2 \,
\delta\left(\left[\frac{\pla E^2}{4}\right],kr^2\right)
\frac{\pla E}{2r^3}\nonumber
\end{eqnarray}
%
%
\begin{equation}\label{density} \end{equation}
where $L$ has been substituted in terms of $R$ and we have defined
\begin{eqnarray}
\Lambda(R) = \left(\frac{R}{\sqrt{\pla}}\,\vartheta(\sqrt{\pla}-R)+
\frac{\sqrt{\pla}}{R}\, \vartheta(R-\sqrt{\pla}) \right)^{17}\\
\Upsilon(R) = \left(\frac{R}{\sqrt{\pla}}\right)^9 + \left(\frac{\sqrt{\pla}}{R}\right)^9 
\end{eqnarray}
As expected, $R$-duality is explicit in the single string density of states through $\Lambda(R)$ and $\Upsilon(R)$ which both encode all the information about the volume dependence.

\section{Thermodynamics: Temperature vs. Energy}

In this section we will study some properties of the single and multiple string gases in this totally compactified universe. We start by showing that the infinite volume limit of the density of states of the system is completely equivalent to the one obtained for an already open universe using the inverse Laplace transform techniques. Those methods were used in\cite{us} for the bosonic string case and can easily be adapted to the Superstring case giving  

\begin{equation}
\label{compom}
\Omega_1(E)=\frac{V\pi^{-9/2}}{\Gamma\left(\frac{9}{2}\right)}
\sum_{(r\geq 1\, ,odd)}\sum_{k=0}^{\infty}r^{-10}a_k^2
E\left(E^2-\frac{4 r^2 k}{\pla}\right)^{7/2} \vartheta \left(E^2-\frac{4 r^2 k}{\pla}\right)
\vspace{4mm}
\end{equation}
with $(V=(2\pi R)^9)\rightarrow \infty$. Taking the limit $R\rightarrow \infty$ in eq.(\ref{density}) it is immediate to verify that both limits exactly coincide.  The last term in eq.(\ref{density}) does not depend on $R$ so that it gives a finite contribution to the infinite volume limit. It does not appear when one uses the $S$-representation for the Helmholtz free energy because the use of continuous momenta reduces the measure of these states in the integral to zero. This is a well known fact in Solid State Physics. These states are equivalent to those that produce Bose condensation (cf. for instance \cite{huang}). In any case they have to be taken into account separately. For an open universe  we can neglect them compared to those that diverge linearly  with the volume.

	Once we have got the density of states we are able to study all the thermodynamical properties of the gas. The temperature is 

\begin{equation}
T^{-1}(E) = \left[\frac{1}{\Omega(E,V)}\frac{\partial\Omega(E,V)}{\partial E}\right]_V 
\label{temperature}
\end{equation}

\begin{figure}
\let\picnaturalsize=N
\def\picsize{2in}
\def\picfilename{single.eps}
\ifx\nopictures Y\else{\ifx\epsfloaded Y\else\input epsf \fi
\let\epsfloaded=Y
\centerline{\ifx\picnaturalsize N\epsfxsize \picsize\fi
\epsfbox{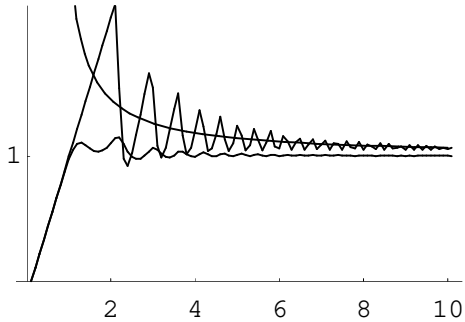}}}\fi
\caption{\small{The temperature of a single closed Superstring gas in terms of its energy.}}
\label{fsingle}
\end{figure}

	In figure\footnote{In our units, the Hagedorn temperature is one and energy is measured in units of  $ 1/\sqrt{\pla}$}\ref{fsingle} we present the temperature calculated for the single string system. The oscillating plot represents the temperature of the string for a big (infinite, for physical purposes) volume and the lower plot corresponds to the same system confined into a small box. Both are compared to the temperature calculated for the asymptotic high energy limit where the density of states is approximated by: 
\begin{equation}
\Omega^{\Upsilon}_ {asymp}\propto\frac{e^{\beta_HE}}{E^{11/2}}\,\Upsilon (R)
\label{asymptotic}
\end{equation}
This is the leading term for big volumes. One can also compute the high energy behaviour of the first term in eq.(\ref{density}) to give

\begin{equation}
\Omega^{\Lambda}_{asymp}\propto\frac{e^{\beta_HE}}{E^{1/2}}\,\Lambda (R)
\end{equation}

\noindent As expected,  at low energies the system behaves like a set of massless quantum fields in the microcanonical description, that is, its specific heat is positive and constant.
Before going on  we should mention that for those low energies and small volume the spectrum gets very discrete and no thermodynamical magnitude can in fact be defined. From the  point of view  of Statistical Mechanics we cannot assume that $\Delta E$ is small because in this case it is as big as the total energy of the system. Increasing the volume or the energy one can recover a thermodynamical description since $\Delta E$ becomes small compared to the energy values.  

For a given volume allowing a thermodynamical description at low energies, the accesible string states are those that are made from either pure momenta or windings, the election is irrelevant because of $R$-duality. These strings do not distinguish small volumes from large ones so that, in this range of energies, neither topology nor size affect the temperature, in such a way that the specific heat is the same in any situation. A purely stringy effect like the existence of states that combine windings and momenta changes this behaviour, cooling the gas. This effect gets less important as the volume grows. 

The oscillations of the temperature are due to the opening of ever more degenerate mass thresholds corresponding to the oscillator energy levels of the string. This explains the successive coolings of the gas\footnote{Any relativistic quantum theory with several discrete mass levels presents this kind of oscillations in the temperature. }. In the plots the maxima look very sharp but the function $T(E)$ has a continuous derivative. Because of the preponderance of the Maxwell-Boltzmann statistics at high energies the system tends to have the Hagedorn temperature.

Now that we have the complete knowledge of the thermal properties of the single string we are able to attempt to study the whole system. The multiple string gas density of states $\Omega(E)$ can be computed using repeated convolutions as we explained in section two. It would be

\begin{equation}
\Omega(E)=\sum_{N=1}^{\infty} \frac{1}{N!}\Omega_1(E)\ast \stackrel{N)}{\cdot\cdot\cdot}
\ast \Omega_1(E).
\label{multiple}
\end{equation}

\begin{figure}
\let\picnaturalsize=N
\def\picsize{2in}
\def\picfilename{infvol.eps}
\ifx\nopictures Y\else{\ifx\epsfloaded Y\else\input epsf \fi
\let\epsfloaded=Y
\centerline{\ifx\picnaturalsize N\epsfxsize \picsize\fi
\epsfbox{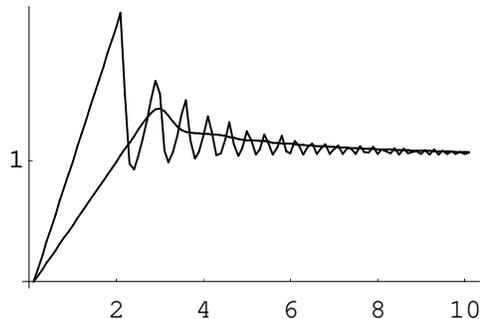}}}\fi
\caption{\small{The temperature of a single and two string gases in an open universe.}}
\label{infvol}
\end{figure}

We have to sum up every contribution because the system is a quantum gas with null chemical potential. In the case of large volume, the contribution of the $N\rightarrow\infty$ term outweighs all the others as $\Omega_1(E)$ grows linearly with the volume. In figure \ref{infvol} we show the temperature of the single and the two string gases in terms of their total energy for that case. The growth of the number of accesible states at a given energy due to the second string, softens the thermal behaviour of the gas, lowering the maximum temperature and supressing the sharp variations that appeared in the single string gas. Extrapolating this evolution to the $N$-string system, it is possible to induce the behaviour with an infinite number of strings. That gives a positive and growing specific heat that diverges at a finite energy per string; there, the system reaches its maximum temperature, the Hagedorn one, and from that point on, the specific heat remains infinite and the temperature constant (see \cite{us}).

\begin{figure}
\let\picnaturalsize=N
\def\picsize{2in}
\def\picfilename{autodual2.eps}
\ifx\nopictures Y\else{\ifx\epsfloaded Y\else\input epsf \fi
\let\epsfloaded=Y
\centerline{\ifx\picnaturalsize N\epsfxsize \picsize\fi
\epsfbox{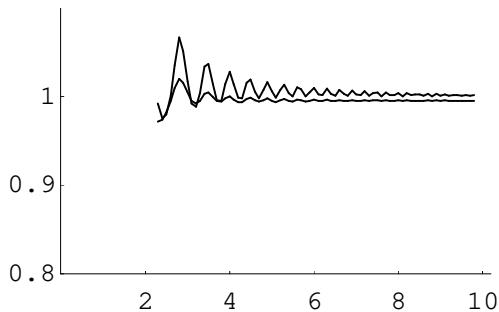}}}\fi
\caption{\small{The temperature of a single and two string gases in a box of selfdual size.}}
\label{autodual}
\end{figure}

The scenario in the finite volume case is quite similar, but there is a finite number of relevant terms in the series (\ref{multiple}) at a given energy. Figure \ref{autodual} shows the temperature for the selfdual volume, in the  energy  range in which the thermodynamical description is defined. The influence of the second term is also the stabilization of the system, just as in the infinite volume case. It can be seen that its relevance at low energies is small and becomes more important at certain energies where it makes the specific heat positive and finite. However if the system reaches the energy at which the $N>2$ terms become relevant, the two-string gas will tend to have a negative specific heat, but precisely at these energies we must sum up the other contributions that stabilize once more the system. In short, the multiple string system has a finite and positive specific heat in the whole energy range, in spite of the fact that every gas with a finite number of strings has got negative specific heat for certain asymptotic ranges of energy. The low energy oscillations that appeared for the single string and that vanish for the multiple string gas are an example of the softening of the thermal behaviour by bigger $N$ terms. The most relevant term of the series is the one that corresponds to the most probable number of strings in the system. For instance, at low energies, this number is one.

\section{Thermodynamics: Pressure and $R$-duality}
\begin{figure}
\let\picnaturalsize=N
\def\picsize{2in}
\def\picfilename{phase.ps}
\ifx\nopictures Y\else{\ifx\epsfloaded Y\else\input epsf \fi
\let\epsfloaded=Y
\centerline{\ifx\picnaturalsize N\epsfxsize \picsize\fi
\epsfbox{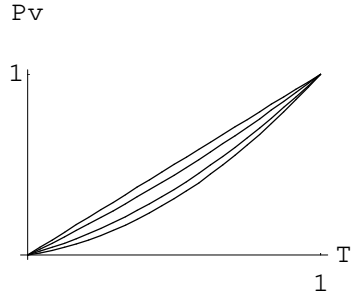}}}\fi
\caption{\small{The phase diagram of the string gas for several volumes larger than the selfdual one. The system reaches a maximum pressure when $T=T_H$, {\tt v} equals $V/N$, where $N$ is the number of strings. }}
\label{phase}
\end{figure}

In the microcanonical ensemble the pressure of a system is usually defined by
\begin{equation}
P(E,V)=\left[T(E,V)\frac{\partial \log(\Omega(E,V))}{\partial V}\right]_E
\end{equation}
If we understand $R$-duality as a symmetry of the system we should expect that  an observer could not distinguish which of the two possible values of $V$ he/she is living in, so any physical magnitude must be invariant under duality. It is evident that the usual definition of pressure does not fit into this criterium. This is why we are in need of a new, dual definition of pressure. There are several possibilities but the one that seems to us the most appropriate in order to keep its usual thermodynamical features is 
\begin{eqnarray}
P(E,V)=T(E,V)\left[\frac{\partial \log(\Omega(E,V))}{\partial V}\,\vartheta(V-1)+\frac{\partial \log(\Omega(E,V))}{\partial (V^{-1})}\,\vartheta(1-V)\right]_E \nonumber\ \\
\label{newp}
\end{eqnarray}
where we express the volume in units of the selfdual one. It is immediate to modify every other thermodynamical relation in which the volume or pressure appear. With this definition, we recover the usual thermodynamics for $V>1$. All these subtleties arise because of a redundant  treatment of the string's phase space as it truly consists only of half the configurations we are considering. That is, the region with volumes smaller than the selfdual one is exactly the same that the $V>1$ one. 
For a given $(V>1)$ volume the system at low energies is such that the states   of pure momentum are filled and so the pressure tends to grow as the energy does. When winding states are accesible they contribute negatively to the pressure, but in this situation equipartition breaks (see \cite{us,DJT}) and the combined states of winding and momentum are in fact only accesible to one string of the total system. The others remain in the low energy states, which are accounted for by the term with $\Upsilon(R)$ in (\ref{density}), and  contribute positively to the pressure. The final result is that the complete gas has a positive pressure and reaches an asymptotic value $(T_H N/ V)$ at very high energies. The case of infinite volume is special because winding states are abstent so there is no negative contribution to the pressure. We see this behaviour in the qualitative $P\mbox{{\tt v}}$ vs. $T$ diagram in the microcanonical ensemble shown in fig \ref{phase}. 

Another interesting feature of the statistical behaviour of the system is that its entropy always grows with the volume, getting a minimum for the selfdual size, at any energy.

\section{Conclusions}
We have studied a microcanonical description of a container of finite volume $V$ filled with strings.
The problem of counting the accessible states at a given energy has been solved indirectly passing through the same counting for a single string filling the whole box. The next step toward the description of the physical system with an average number of strings has been done by using repeated convolutions.

The single string degeneration has been calculated  using  Condensed Matter standard techniques in the sense that we use the mass formula  and the constraint as  a dispersion relation that gives the energy as a function of the quantum numbers defining the states. So finally the task is reduced to a problem of computing volumes in phase space.

Gotten the density of states, it is easy to compute the infinite volume limit degeneration to get exactly the same result one obtains by inverse Laplace transforming $(\log{Z(\beta)})$ for an already open universe. The information about the multiple string system is actually seeded in the single string density and is gotten by simply  putting a set of strings to share a given amount of energy. This is roughly speaking what amounts to  making a convolution.

Another result is that, at finite volume, whenever there is enough states to make statistics, the specific heat is positive and finite and the Hagedorn temperature is reached asymptotically. More precisely, the term that combines winding and momenta and unbalanced  left and right oscillators $(N \neq \tilde{N})$ makes the difference at finite volume. At low energies and small  volumes the discrete nature of the spectrum prevents any  thermodynamical description although one can say that there is only a single string at a fixed energy.

In this work  we assume as a belief that $R$-duality is a symmetry like gauge symmetry in the sense that there is a redundancy in phase space  resulting from the assumed complete equivalence between the system at size $R$ and $\pla/R$. Then we propose that the physical sizes go either from the selfdual point to infinity or from the selfdual point to zero. This point of view has important consequences because no physical measure may elucidate between the two pictures.
This interpretation is clearly incompatible with the pressure as it is usually defined. So, we propose a new definition compatible with $R$-duality which reduces to the standard one for volumes bigger than the selfdual one. This way the dependence on the volume of all the thermodynamical magnitudes is smooth.

One of the most remarkable things is that the entropy of the multiple  string gas depends on the volume in such a way that there is an absolute minimum at the selfdual size and grows with the volume in any other case. If the equilibrium states are those with maximum entropy, then our finite size box should expand to maximize the degeneration and finally reach thermodynamical equilibrium.

\newpage
\section*{Acknowledgements}
We thank M.A.Gonz\'{a}lez, J. Junquera and L. Blanco and other people of the Fundamental Physics Group in Oviedo for encouragement and  useless but nice discussions. Marco and Jes\'{u}s are still in debt with their parents for economic support.

\end{document}